\documentclass[twocolumn,showpacs,preprintnumbers,amsmath,amssymb]{revtex4}
\usepackage{graphicx}% Include figure files
\usepackage{dcolumn}% Align table columns on decimal point
\usepackage{bm}% bold math

\begin{document}
\title{Physical properties in hole-doped SrFe$_{2-x}$Cu$_x$As$_2$ single crystals}

\author{Y. J. Yan, P. Cheng, J. J. Ying, X. G. Luo, F. Chen, H. Y. Zou, A. F. Wang, G. J. Ye, Z. J. Xiang, J. Q. Ma,}
\author{X. H. Chen$^1$}
\altaffiliation{Corresponding author} \email{chenxh@ustc.edu.cn}
\affiliation{Hefei National Laboratory for Physical Science at
Microscale and Department of Physics, University of Science and
Technology of China, Hefei, Anhui 230026, People's Republic of
China}

\date{\today}% It is always \today, today,

\begin{abstract}
We report the structural, magnetic and electronic transport
properties of SrFe$_{2-x}$Cu$_x$As$_2$ single crystals grown by
self-flux technique. SrCu$_2$As$_2$ and SrFe$_2$As$_2$ both
crystallize in ThCr$_2$Si$_2$-type (122-type) structure at room
temperature, but exhibit distinct magnetic and electronic transport
properties. The x-ray photoelectron spectroscopy(XPS) Cu-2\emph{p}
core line position, resistivity, susceptibility and positive Hall
coefficient indicate that SrCu$_2$As$_2$ is an \emph{sp}-band metal
with Cu in the 3\emph{d$^{10}$} electronic configuration
corresponding to the valence state Cu$^{1+}$. The almost unchanged
Cu-2\emph{p} core line position in SrFe$_{2-x}$Cu$_x$As$_2$ compared
with SrCu$_2$As$_2$ indicates that partial Cu substitutions for Fe
in SrFe$_2$As$_2$ may result in hole doping rather than the expected
electron doping. No superconductivity is induced by Cu substitution
on Fe sites, even though the structural/spin density wave(SDW)
transition is gradually suppressed with increasing Cu doping.
\end{abstract}

\maketitle

\section{\label{sec:level1}Introduction}

Exploration of new high-temperature superconductors and research on
their superconducting mechanism have always been the highlight in
condensed matter physics. Especially the discovery of a second class
of high-temperature superconductors - iron-based superconductors,
has reignited the boom of the high-temperature superconductivity
research. There are five types of iron-based superconductors: 1111
with ZrCuSiAs-type structure, 122 with ThCr$_2$Si$_2$-type
structure, 111 with Fe$_2$As-type structure, 11 with anti-PbO-type
structure, and a newly discovered K$_x$Fe$_{2-y}$Se$_2$ with Fe
vacancy which is called 122* structure. Among these materials,
compound with 122 structure has gained much attention because of the
high superconducting temperature and easiness to obtain large high
quality single crystals.

In the 122-type compound BaFe$_2$As$_2$, superconductivity can be
induced by applying pressure\cite{Yamazaki, Ishikawa, Colombier},and
by substitutions at the Ba site (by K)\cite{Chen}, at the Fe site
(by Co, Ni, Ru, Rh, and Pd)\cite{Johnston, Canfield}, and at the As
site (by P)\cite{Jiang}. It has been revealed that partial Co, Ni,
Rh and Pd substitutions at the Fe site in BaFe$_2$As$_2$ could
induce superconductivity\cite{Sefat, Wang, X. F. Wang, Li} with Tc
up to 25 K whereas no superconductivity is induced by
Mn\cite{Kasinathan, Liu, J. S. Kim} or Cr \cite{A. S. Sefat, Marty}
substitutions. As we know, the formal valence states of the atoms in
BaFe$_2$As$_2$ are assigned as Ba$^{2+}$, Fe$^{2+}$ and As$^{3-}$,
so the Fe atoms are formally in the 3\emph{d$^{6}$} electronic
configuration.  So it means that while electron doping at the Fe
sites by Co, Ni with more 3\emph{d} electrons induces
superconductivity, hole doping by Cr, Mn with less 3\emph{d}
electrons does not. From this point of view, divalent copper
Cu$^{2+}$ with three more \emph{d}-electrons than Fe should be a
strong electron dopant for iron arsenide superconductors. However,
even though the Cu doping successfully suppresses the structural/SDW
transition of the parent compound\cite{Ni}, superconductivity was
not observed in the Cu-doped BaFe$_2$As$_2$. It is a very strange
phenomenon that doping Cu into iron-based superconductors shows a
completely different character from that of doping with Co and Ni.
Electronic structure calculations for SrCu$_{2}$As$_{2}$ and
BaCu$_{2}$As$_{2}$ by Singh\cite{Singh} predicted that these
compounds might be \emph{sp}-band metals with the Cu atoms having a
formal valence state of Cu$^{1+}$ and a nonmagnetic and chemically
inert 3\emph{d$^{10}$} electronic configuration. Therefore, we
speculate that the distinct valence state of Cu ions may be the main
reason for the different behavior compared with other transition
metal doped condition.

To obtain insights into the nature of the puzzling properties of
Cu-doped iron-based materials, we have synthesized a series of
SrFe$_{2-x}$Cu$_x$As$_2$ single crystals and investigated their
valence state of copper ions by XPS measurement, and their
structural, magnetic and electronic transport properties.

\section{Materials and Methods}

A series of SrFe$_{2-x}$Cu$_x$As$_2$ single crystals, from
SrFe$_2$As$_2$ (x=0) to SrCu$_2$As$_2$ (x=2.0) were grown by
self-flux technique using high purity Sr, Cu, Fe and As. Prereacted
CuAs and FeAs were used as flux. For the growth of
SrFe$_{2-x}$Cu$_x$As$_2$, the Sr and FeAs, CuAs flux with a molar
ratio of 1:2.5(2-x):2.5x were placed into alumina crucibles and
sealed inside evacuated quartz tubes. The crystal growth was carried
out by heating the samples to 1150 $^{\circ}$C, holding there for 24
h and then cooling to 850 $^{\circ}$C at a rate of 2 $^{\circ}$C/h.
The sizes of the obtained single crystals for 0$\leq$ x $<$ 1.0 were
typically 5 $\times$ 3 $\times$ 0.15 mm$^{3}$, while that for 1.0
$\leq$ x $\leq$ 2.0 were typically 2.5 $\times$ 2 $\times$ 0.15
mm$^{3}$.

The samples were characterized by x-ray diffraction (XRD) using
Rigaku D/max-A x-ray diffractometer with Cu K$_\alpha$ radiation in
the range of 10$^{\circ}$-70$^{\circ}$ with the step of
0.01$^{\circ}$ at room temperature. The actual Cu and Fe
concentration of the single crystals were determined from
energy-dispersive x-ray (EDX) analysis. The Cu content x used in
this article is the actual composition determined by EDX. The
valence state of Cu is determined by x-ray photoelectron
spectroscopy(XPS). The resistivity was measured using the standard
four-probe method by Quantum Design Physical Property Measurement
System (PPMS). Hall resistivity data were collected using the ac
transport option of a quantum design physical property measurement
system in a four-wire geometry with switching the polarity of the
magnetic field H $\parallel$ c to remove any magnetoresistive
components due to the misalignment of the voltage contacts. Magnetic
susceptibility was measured using Vibrating Sample Magnetometer
(VSM).

\section{Results and Discussion}

Figure 1(a) shows the single crystal XRD patterns for all the
SrFe$_{2-x}$Cu$_x$As$_2$ single crystals, and only (00l) diffraction
peaks are observed, indicating that the single crystals are in
perfect (001) orientation. Fig. 1(b) shows the evolution of the
lattice parameters of a- and c-axis as a function of Cu doping
content. The lattice parameter of c-axes is obtained from the (00l)
diffraction peaks, while that of a-axes is obtained by powder XRD.
With increasing Cu doping content, the lattice parameter of a-axes
increases monotonically, while that of c-axes decreases
monotonically. The unit cell volume \emph{V} $=$ a$^{2}$c firstly
increases with increasing Cu content, reaches a maximum at about
x$\sim$1.0, and then decreases with further increasing Cu content.
Obviously, the evolution of \emph{V} for SrFe$_{2-x}$Cu$_x$As$_2$
shows a qualitative deviation from Vegard's law, which shows a
linear decrease with Cu content. Ni \emph{et al}.\cite{Ni} reported
the lattice parameters a and c and the unit cell volume \emph{V} of
Ba(Fe$_{1-x}$Cu$_x$)$_2$As$_2$ versus x up to x$=$0.35, which shows
a similar evolution of \emph{V} to SrFe$_{2-x}$Cu$_x$As$_2$.
Singh\cite{Singh} gave theoretical prediction that Cu in
BaCu$_2$As$_2$ and SrCu$_2$As$_2$ has a fully occupied stable
\emph{d$^{10}$} shell at high binding energy, which means that the
valence state of Cu in these compounds is +1. While in
SrFe$_2$As$_2$, the valence state of Fe is +2. Therefore, we
conclude that the anomalous behavior of \emph{V} of
SrFe$_{2-x}$Cu$_x$As$_2$ may indicate interesting changes of valence
state in the Cu/Fe sites. In order to confirm it, we performed XPS
measurement on several samples, and the results are shown in Fig. 2.
As shown in Fig. 2, the Cu2$\emph{p}_{1/2}$ and Cu2$\emph{p}_{3/2}$
binding energy maxima for Cu metal are about 951.9 eV and 932.0 eV,
and the linewidths are very narrow. The lines in
SrFe$_{2-x}$Cu$_x$As$_2$ single crystals are a little broader than
that in Cu metal, and have positions nearly the same as that in Cu
metal, which is the condition of Cu$_2$O as reported
previously\cite{Steiner}. While for CuO, the lines for divalent
copper are distinctly different. Compared with that of Cu metal, the
lines in CuO shift to higher binding energies, the linewidth is
broader with a factor of 2, and also the satellite is very intense.
Considering these distinct differences, the valence of Cu ions in
SrFe$_{2-x}$Cu$_x$As$_2$ single crystals is monovalent, regardless
of doping level.

\begin{figure}[ht]
\centering
\includegraphics[width=0.68\textwidth]{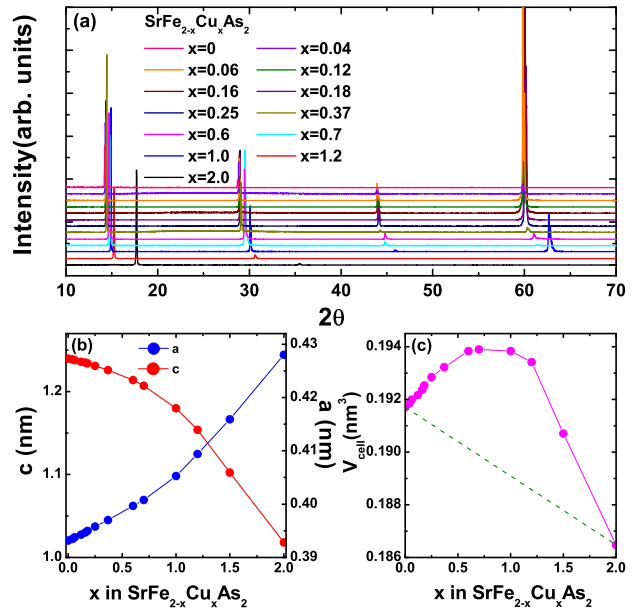}
\caption{(Color online)(a): Single crystal x-ray diffraction
patterns at room temperature for SrFe$_{2-x}$Cu$_x$As$_2$ single
crystals (x is the actual composition covering from 0 to 2.0.). Only
(00l) diffraction peaks are observed, indicating that the c axis is
perpendicular to the plane of the single crystal. (b): Lattice
parameters of a- and c-axis as a function of x. The lattice
parameters of a- and c-axis were obtained by combining single
crystal XRD and powder XRD patterns. The data for x=1.5 are
collected from polycrystal. (c): The unit cell volume \emph{V} as a
function of x for SrFe$_{2-x}$Cu$_x$As$_2$ single crystals. The
green dashed line represent Vegard's law for this series of
compounds.}\label{Fig1}
\end{figure}

\begin{figure}[ht]
\centering
\includegraphics[width=0.55\textwidth]{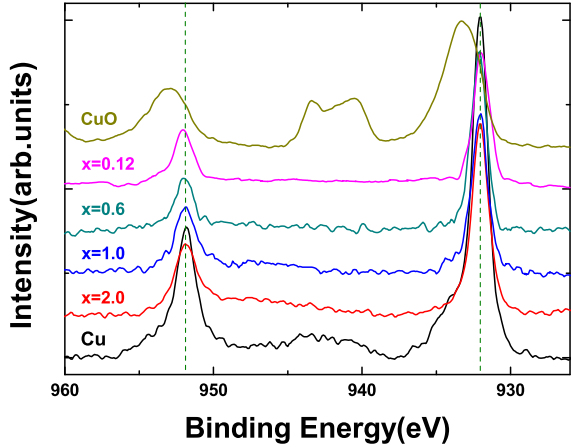}
\caption{(Color online) XPS Cu - 2$\emph{p}$ spectra of Cu metal
(where Cu is nominally in a Cu$^{0+}$ state), CuO (where Cu is in a
Cu$^{2+}$ state) and SrFe$_{2-x}$Cu$_x$As$_2$ single crystals with
x=0.12, 0.6, 1.0, 2.0 after background subtraction. The two dashed
lines correspond to the binding energies of Cu - 2$\emph{p}$$_{1/2}$
and Cu - 2$\emph{p}$$_{3/2}$ in Cu metal. }\label{Fig2}
\end{figure}

The temperature dependence of in-plane electrical resistivity for
SrFe$_{2-x}$Cu$_x$As$_2$ single crystals are shown in Fig. 3. As
shown in Fig. 3(a), the resistivity exhibits obvious anomaly at the
temperature from 196 K for SrFe$_2$As$_2$ to 71 K for the samples
slightly doped with Cu, which is associated with structural/SDW
transition reported previously\cite{Jesche}. The temperature of
structural/SDW transition, \emph{T$_{SDW}$}, decreases with
increasing Cu doping, and disappears when x $>$ 0.25. The anomaly
due to the structural/SDW transition becomes more pronounced with
doping, which is similar to that of Mn or Cr doped
BaFe$_2$As$_2$\cite{J. S. Kim, Sefat}, and in strong contrast with
Co or Ni doped BaFe$_2$As$_2$\cite{Canfield}. Fig. 3(b) shows the
temperature dependence of resistivity for 0.37 $\leq$ x $\leq$ 1.2
samples. The value of resistivity decreases quickly with increasing
Cu doping, and the behavior of resistivity evolves from
semiconducting to metallic-like and no anomaly is observed due to
the structural or magnetic transition. For the samples x=0.37, 0.6,
0.7, the temperature dependence of resistivity shows semiconducting
behavior in the whole temperature range. While for x=1.0, the
resistivity exhibits metallic behavior above 190 K, and then turns
into semiconducting behavior. For x=1.2, the resistivity is 0.19
m$\Omega$ cm at 300 K, which is one order of magnitude larger than
that of SrCu$_2$As$_2$. The resistivity decreases with cooling,
reaches a minimum at around 20 K, and then turns upward. The
resistivity of SrCu$_2$As$_2$ as a function of temperature is
presented in Fig. 3(c), which is similar to the results reported
previously\cite{Anand}. The temperature coefficient of
\emph{$\rho$}(\emph{T}) is positive, indicating metallic character.
The values of residual resistivity \emph{$\rho$$_0$} and residual
resistivity ratio(RRR) are 4.2 $\mu\Omega$ cm and 6.4, respectively,
indicating good quality of SrCu$_2$As$_2$ single crystals. The clear
evolution of the resistivity behavior and phase transition are shown
in Figs. 3(d) and 3(e).

\begin{figure}[ht]
\centering
\includegraphics[width=0.55\textwidth]{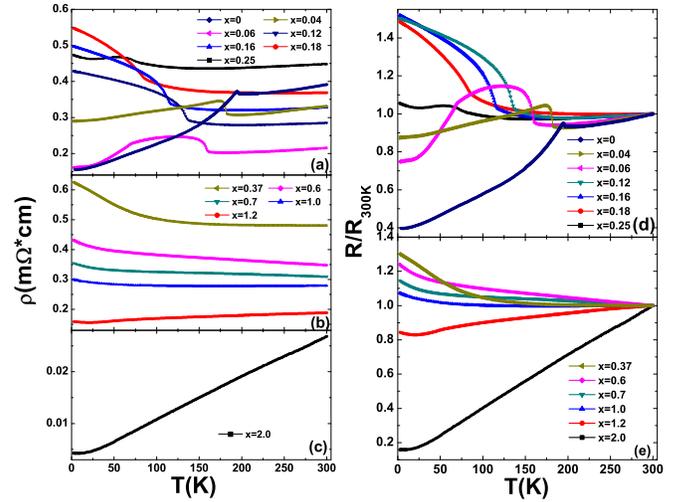}
\caption{(Color online) (a)-(c):Temperature dependence of in-plane
electrical resistivity for SrFe$_{2-x}$Cu$_x$As$_2$ single crystals.
(d) and (e): Temperature dependence of the normalized resistivity to
300 K. }\label{Fig3}
\end{figure}

\begin{figure}[ht]
\centering
\includegraphics[width=0.55\textwidth]{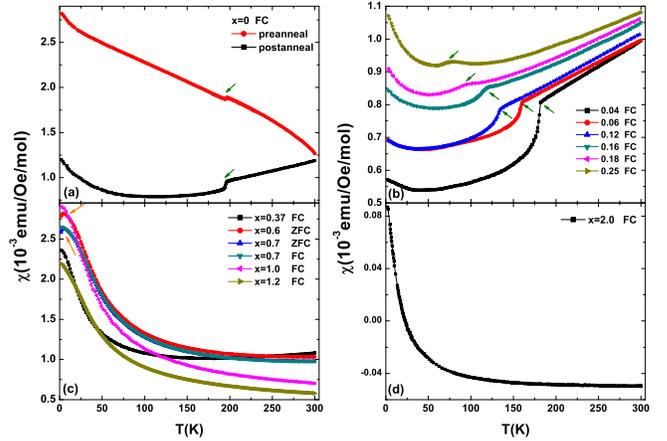}
\caption{(Color online) Temperature dependence of magnetic
susceptibility for SrFe$_{2-x}$Cu$_x$As$_2$ (0 $\leq$ x $\leq$ 2.0)
under magnetic field of 5T. The SDW transition temperature
\emph{$T_{SDW}$} is shown by green arrows. The orange arrow shows
the spin glass-like transition.}\label{Fig4}
\end{figure}

Figure 4 presents the zero-field-cooled (ZFC) or field-cooled(FC)
magnetic susceptibility \emph{$\chi$} for all the single crystals as
a function of temperature from 2 K to 300 K in an applied magnetic
field \emph{H}$=$5.0 T aligned in the ab plane. Fig. 4(a) presents
the typical behavior of one SrFe$_{2}$As$_{2}$ single crystal before
and after an annealing at 300$^{\circ}$C for 5 hours. The
temperature dependence of \emph{$\chi$} exhibits an unusual behavior
for the as-grown SrFe$_{2}$As$_{2}$, and changes to a universal
behavior after annealing, which is consistent with the previous
report\cite{Yan}. These phenomena are supposed to be related to the
presence of lattice distortion in SrFe$_{2}$As$_{2}$\cite{Saha}. For
samples with 0 $\leq$ x $\leq$ 0.25 as shown in Figs. 4(a) and 4(b),
structural/SDW transition is obviously observed and marked by green
arrows, corresponding to the anomaly observed in the resistivity
shown in Figs. 3(a) and 3(d). The \emph{$\chi$}(\emph{T}) of samples
with 0.37 $\leq$ x $\leq$ 1.2 show paramagnetic behavior in the
whole temperature range, and no obvious magnetic transition was
observed down to 2 K. The magnitude of \emph{$\chi$} decreases with
increasing Cu doping as shown in Fig. 4(c). Below 20 K, a small
separation between FC and ZFC curves for the crystal with x=0.7 is
observed, indicating glass-like behavior. The \emph{$\chi$} of
SrCu$_2$As$_2$ in Fig. 4(d) has a negative sign above 20 K, which is
a typical behavior for nonmagnetic metal and in consistent with the
previous report\cite{Anand}.

\begin{figure}[ht]
\centering
\includegraphics[width=0.70\textwidth]{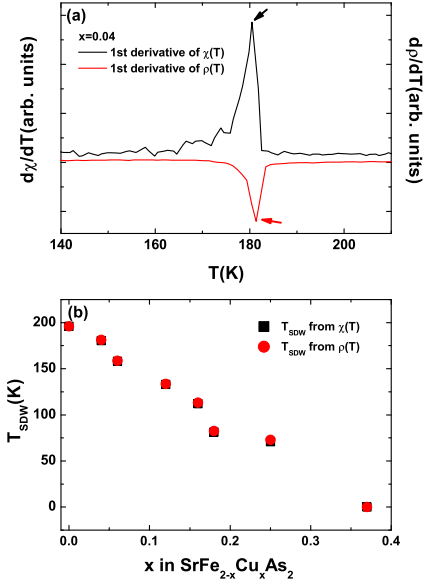}
\caption{(Color online)(a): The derivatives of \emph{$\chi$(T)} and
\emph{$\rho$(T)} as a function of temperature for x$=$0.04 sample.
The distinct peaks indexed by arrows in d\emph{$\chi$(T)}/d\emph{T}
and d\emph{$\rho$(T)}/d\emph{T} curves are used to determine the
temperature of structural$/$SDW transition. (b): Evolution of
\emph{T$_{SDW}$} with Cu doping.}\label{Fig5}
\end{figure}

Figure 5(a) shows the typical derivatives of \emph{$\chi$}(\emph{T})
and \emph{$\rho$}(\emph{T}) to figure out the transition temperature
\emph{T$_{SDW}$} of structural$/$SDW transition. Only one obvious
peak is observed in both derivative curves of \emph{$\chi$(T)} and
\emph{$\rho$(T)}, corresponding to the temperature of phase
transition. \emph{T$_{SDW}$} determined from derivatives of
\emph{$\chi$(T)} and \emph{$\rho$(T)} are highly consistent with
each other, as shown in Fig. 5(b). With increasing Cu doping,
\emph{T$_{SDW}$} decreases quickly from 196 K for x=0 to 71 K for
x=0.25, which is similar to that of
Ba(Fe$_{1-x}$Cu$_x$)$_2$As$_2$\cite{Ni}, except that no
superconductivity was observed in SrFe$_{2-x}$Cu$_x$As$_2$ for any
doping content.

\begin{figure}[ht]
\centering
\includegraphics[width=0.55\textwidth]{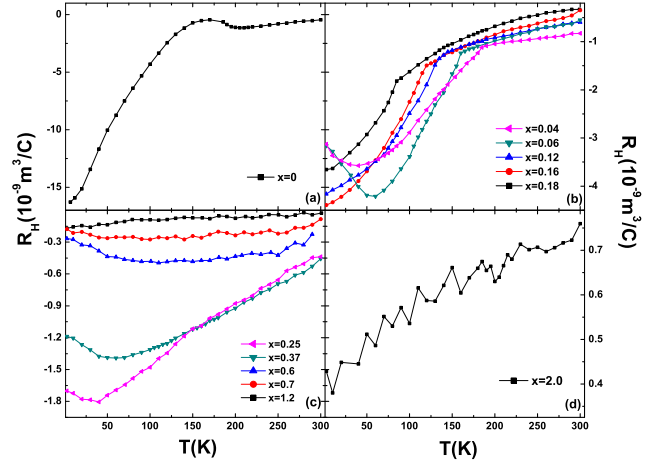}
\caption{(Color online)Temperature dependence of Hall coefficients
\emph{R$_H$} for SrFe$_{2-x}$Cu$_x$As$_2$ samples.}\label{Fig6}
\end{figure}

To further understand the conducting carriers in the Cu-doped
SrFe$_2$As$_2$ samples, the Hall coefficient(\emph{R$_H$})
measurements were carried out on the single crystals, and the
obtained results are shown in Fig. 6. In Figs. 6(a) and 6(b),
distinct structural/SDW transition can be observed for samples 0
$\leq$ x $\leq$ 0.18, consistent with the resistivity and
susceptibility. The Hall coefficient is negative indicating
electron-type carriers dominate, and the absolute value decreases
with Cu doping content. For 0.25 $\leq$ x $\leq$ 1.2, the Hall
coefficients \emph{R$_H$} are also negative in the whole temperature
range, indicating that the dominated carrier is electron-type. The
absolute value of Hall coefficients for x=0.25 and 0.37 are smaller
than that of samples with lower Cu doping content. It increases with
cooling, reach minima at around 40 K, and then weakly decline at low
temperature. For x=0.25, no phase transition is observed around the
temperature corresponding to structural/SDW transition, probably
because its phase transition is too weak. A weak temperature
dependence of Hall coefficient is observed in the whole temperature
region for samples x=0.6, 0.7 and 1.2. The absolute values of
\emph{R$_H$} for these three samples are very small with the same
order of magnitude of 10$^{-10}$ m$^{3}$$/$C, and decreases with
further increasing Cu doping. As shown in Fig. 6(d), it is clear
that the Hall coefficient of SrCu$_2$As$_2$ is positive in the whole
temperature range, which suggests that the hole-type carriers
dominate. The value of \emph{R$_H$} is 7 $\times$ 10$^{-10}$
m$^{3}$$/$C at 300 K, which is remarkably small, indicating a
relatively high density of charge carriers estimated to be of the
order of 10$^{22}$ cm$^{-3}$. The magnitude and temperature
dependence of Hall coefficient for SrCu$_2$As$_2$ is similar to the
reports by Qin et al.\cite{Qin}. In conclusion, with the gradual
increase of Cu doping, the Hall coefficient changes from negative to
positive, and its absolute value changes gradually. These results
indicate that the carrier changes from electron-type to hole-type,
and the concentration of carrier changes simultaneously with Cu
content. Combining with XPS results, it strongly suggests that
partial Cu substitutions for Fe in SrFe$_2$As$_2$ may result in hole
doping rather than the expected electron doping. This is probably
the reason that Cu substitutions in iron-based materials can not
induce superconductivity.

To our knowledge, the valence state of Cu found in the previously
reported pnictide oxides is always +1, because the anionic
environment of the pnictide oxides is in general not sufficiently
electronegative to oxidize Cu to +2. Even in the LaNiO$_2$-type
[M'O$_2$] layer of oxysulfides
Sr$_2$[M'$_{1-x}$Cu$^{2+}$$_x$O$_2$][Cu$^{1+}$$_2$S$_2$] (M' = Sc,
Cr, Mn, Fe, Co, Ni, Zn), which are expected to have more
electronegative anionic environment than the pnictide oxides, the
maximum Cu$^{2+}$ content for the single-phase sample was x $<$ 1
\cite{Okada, Hirose}. This also indicates that the accommodation of
Cu$^{2+}$ in the weakly electronegative anionic environment of
suboxides such as pnictide oxides and oxychalcogenides is difficult
in contrast to its accommodation in the more electronegative anionic
environment of oxides\cite{Tadashi}. Therefore, the valence state of
Cu prefers +1 in the iron-based pnictide with more weakly
electronegative anionic environment, such as in EuCuPn(Pn = P, As,
Sb)\cite{Michel}, BaCuAs, CaCuAs, SrCu$_{2}$As$_{2}$,
SrCu$_{2}$Sb$_{2}$, and BaCu$_{2}$Sb$_{2}$\cite{Anand}.

\section{Conclusion}

In summary, XPS Cu-2$\emph{p}$ spectra, deviation of \emph{V} from
Vegard's law, diamagnetic susceptibility, and positive Hall
coefficient indicate that the valence state of Cu in SrCu$_2$As$_2$
is +1 with a fully occupied 3\emph{d} shell. The structural/SDW
transition is suppressed with increasing Cu doping content for x
$\leq$ 0.25, and disappears when Cu doping content x is higher than
0.25. Superconductivity can not be induced over the whole doping
range. The nearly same Cu-2\emph{p} core line position as that in
SrCu$_2$As$_2$ for all crystals and the evolution of Hall
coefficients strongly suggest that partial Cu substitutions for Fe
in SrFe$_2$As$_2$ may result in hole doping rather than the expected
electron doping, which is the possible reason for that Cu doping
cannot induce superconductivity in the SrFe$_{2-x}$Cu$_x$As$_2$.

\vspace*{2mm} {\bf Acknowledgment:} This work is supported by the
National Natural Science Foundation of China, and the National Basic
Research Program of China (973 Program, Grants No. 2012CB922002 and
No. 2011CB00101), and the Chinese Academy of Sciences.


\begin{thebibliography}{}
\bibitem{Yamazaki}
T. Yamazaki, N. Takeshita, R. Kobayashi, H. Fukazawa, Y. Kohori, K.
Kihou, C. H. Lee, H. Kito, A. Iyo, and H. Eisaki, Phys. Rev. B
\textbf{81}, 224511 (2010).
\bibitem{Ishikawa}
F. Ishikawa, N. Eguchi, M. Kodama, K. Fujimaki, M. Einaga, A.
Ohmura, A. Nakayama, A. Mitsuda, and Y. Yamada, Phys. Rev. B
\textbf{79}, 172506 (2009).
\bibitem{Colombier}
E. Colombier, S. L. Bud¡¯ko, N. Ni, and P. C. Canfield, Phys. Rev. B
\textbf{79}, 224518 (2009).
\bibitem{Chen}
H. Chen, Y. Ren, Y. Qiu, Wei Bao, R. H. Liu, G. Wu, T. Wu, Y. L.
Xie, X. F. Wang, Q. Huang, X. H. Chen, Europhys. Lett. \textbf{85},
17006 (2009).
\bibitem{Johnston}
D. C. Johnston, Adv. Phys. \textbf{59}, 803 (2010).
\bibitem{Canfield}
P. C. Canfield, and S. L. Bud¡¯ko, Annu. Rev. Condens. Matter Phys.
\textbf{1}, 27 (2010).
\bibitem{Jiang}
S. Jiang, H. Xing, G. Xuan, C. Z. Ren, C. Feng, J. Dai, Z. Xu, and
G. Cao, J. Phys. Condens. Matt. \textbf{21}, 382203 (2009).
\bibitem{Sefat}
A. S. Sefat, R. Jin, M. A. McGuire, B. C. Sales, D. J. Singh, and D.
Mandrus, Phys. Rev. Lett. \textbf{101}, 117004 (2008).
\bibitem{Wang}
C. Wang, Y. K. Li, Z. W. Zhu, S. Jiang, X. Lin, Y. K. Luo, S. Chi,
L. J. Li, Z. Ren, M. He, H. Chen, Y. T. Wang, Q. Tao, G. H. Cao, and
Z. A. Xu, Phys. Rev. B \textbf{79}, 054521 (2009).
\bibitem{X. F. Wang}
X. F. Wang, T. Wu, G. Wu, R. H. Liu, H. Chen, Y. L. Xie and X. H.
Chen, New J. Phys. \textbf{11}, 045003 (2009).
\bibitem{Li}
L. J. Li, Q. B. Wang, Y. K. Luo, H. Chen, Q. Tao, Y. K. Li, X. Lin,
M. He, Z. W. Zhu, G. H. Cao, and Z. A. Xu, New J. Phys. \textbf{11},
025008 (2009).
\bibitem{Kasinathan}
D. Kasinathan, A. Ormeci, K. Koch, U. Burkhardt, W. Schnelle, A.
Leithe-Jasper, and H. Rosner, New J. Phys. \textbf{11}, 025023
(2009).
\bibitem{Liu}
Y. Liu, D. L. Sun, J. T. Park, and C. T. Lin, Physica C
\textbf{470}, S513 (2010).
\bibitem{J. S. Kim}
J. S. Kim, S. Khim, H. J. Kim, M. J. Eom, J. M. Law, R. K. Kremer,
J. H. Shim, and K. H. Kim, Phys. Rev. B \textbf{82}, 024510 (2010).
\bibitem{A. S. Sefat}
A. S. Sefat, D. J. Singh, L. H. VanBebber, Y. Mozharivskyj, M. A.
McGuire, R. Jin, B. C. Sales, V. Keppens, and D. Mandrus, Phys. Rev.
B \textbf{79},224524 (2009).
\bibitem{Marty}
K. Marty, A. D. Christianson, C. H. Wang, M. Matsuda, H. Cao, L. H.
VanBebber, J. L. Zarestky, D. J. Singh, A. S. Sefat, and M. D.
Lumsden, Phys. Rev. B \textbf{83},060509(R) (2011).
\bibitem{Ni}
N. Ni, A. Thaler, J. Q. Yan, A. Kracher, E. Colombier, S. L. Bud'ko,
P. C. Canfield, and S. T. Hannahs, Phys. Rev. B \textbf{82}, 024519
(2010).
\bibitem{Singh}
D. J. Singh, Phys. Rev. B \textbf{79}, 153102 (2009).
\bibitem{Steiner}
P. Steiner, V. Kinsinger, I. Sander, B. Siegwart, S. H\"{u}fner, C.
Politis, R. Hoppe, and H. P. M\"{u}ller, Z. Phys. B - Condensed
Matt. \textbf{67}, 497 (1987).
\bibitem{Jesche}
A. Jesche, N. Caroca-Canales, H. Rosner, H. Borrmann, A. Ormeci, D.
Kasinathan, H. H. Klauss, H. Luetkens, R. Khasanov, A. Amato, A.
Hoser, K. Kaneko, C. Krellner,and C. Geibel, Phys. Rev. B
\textbf{78}, 180504(R) (2008).
\bibitem{Anand}
V. K. Anand, P. Kanchana Perera, Abhishek Pandey, R. J. Goetsch, A.
Kreyssig, and D. C. Johnston, Phys. Rev. B \textbf{85}, 214523
(2012).
\bibitem{Yan}
J.-Q. Yan, A. Kreyssig, S. Nandi, N. Ni, S. L. Bud¡¯ko, A. Kracher,
R. J. McQueeney, R. W. McCallum, T. A. Lograsso, A. I. Goldman, and
P. C. Canfield1, Phys. Rev. B \textbf{78}, 024516 (2008).
\bibitem{Saha}
S. R. Saha, N. P. Butch, K. Kirshenbaum, and Johnpierre Paglione,
and P.Y. Zavalij, Phys. Rev. Lett. \textbf{103}, 037005 (2009).
\bibitem{Qin}
Mingsheng Qin, Chongyin Yang, Yaoming Wang, Zhongtian Yang, Ping
Chen, and Fuqiang Huang, J. Solid State Chem. \textbf{187}, 323
(2012).
\bibitem{Okada}
S. Okada,M. Matoba, s. Fukumoto, S. Soyano, Y. Kamihara, T.
Takeuchi, H. Yoshida, K. Ohoyama, and Y. Yamaguchi, J. Appl. Phys.
\textbf{91}, 8861 (2002).
\bibitem{Hirose}
H. Hirose, K. Ueda, H. Kawazoe, and H. Hosono, Chem. Mater.
\textbf{14}, 1037 (2002).
\bibitem{Tadashi}
T. C. Ozawa, and S. M. Kauzlarich, Sci. Technol. Adv. Mater.
\textbf{9}, 033003 (2008).
\bibitem{Michel}
G. Michels, S. Junk. W. Schlabitz, E. Holland-Moritz, M. M.
AbdElmeguidt, J. D\"{u}nner, and A Mewis, J. Phys.: Condens. Matter
\textbf{6}, 1769 (1994).
\end{thebibliography}
\end{document}